\documentclass[journal]{IEEEtran}
\usepackage{textcomp}
\usepackage{xcolor}
\usepackage{graphicx}
\usepackage{amsmath}
\usepackage{amssymb}
\usepackage{amsfonts}
\usepackage{epsfig}
\usepackage{epstopdf}
\usepackage{hhline}
\usepackage{stfloats}
\usepackage{cite}

\usepackage[CJKbookmarks=true,
bookmarksnumbered=true,
bookmarksopen=true]{hyperref}
\usepackage{multirow}
\usepackage{subfigure}

\usepackage{algorithm}
\usepackage{algorithmic}

\usepackage{booktabs}

\DeclareMathOperator*{\argmin}{arg\,min}

\begin{document}

\title{Large and Small Model Collaboration for \\Air Interface}

\author{Yiming~Cui, \emph{Graduate Student Member}, \emph{IEEE},
Jiajia~Guo, \emph{Member}, \emph{IEEE},
Xiao~Li, \emph{Member}, \emph{IEEE},
Chao-Kai~Wen, \emph{Fellow}, \emph{IEEE}, and Shi~Jin, \emph{Fellow}, \emph{IEEE}

\thanks{
Y. Cui, J. Guo, X. Li, and S. Jin are with the School of Information Science and Engineering, Southeast University, Nanjing 210096, China (e-mail: cuiyiming@seu.edu.cn; jiajiaguo@seu.edu.cn; li\_xiao@seu.edu.cn; jinshi@seu.edu.cn).

C.-K. Wen is with the Institute of Communications Engineering, National Sun Yat-sen University, Kaohsiung 80424, Taiwan (e-mail: chaokai.wen@mail.nsysu.edu.tw).
}

\thanks{An earlier version of this paper was submitted to the IEEE ICC 2026 \cite{conference}.}
}	

\maketitle
\thispagestyle{empty}

\begin{abstract}
Large artificial intelligence models (LAMs) have shown strong capability in wireless communications, yet existing works mainly rely on their generalized knowledge across environments while overlooking the potential gains of environment-specific adaptation. Directly fine-tuning LAMs for adaptation is often impractical due to prohibitive training costs, low inference efficiency in multi-user scenarios, and the risk of catastrophic forgetting, in addition to the limited accessibility of model parameters. 
To address these limitations, we establish a collaborative framework for air interface. In this framework, unlike prior approaches that either depend solely on LAMs or require direct fine-tuning, LAMs are exploited as a universal channel knowledge base while small artificial intelligence models (SAMs) are employed as lightweight plugins to capture environment-specific knowledge, facilitating efficient environment-specific adaptation of LAMs.
Subsequently, we instantiate this framework for CSI feedback tasks, and develop a \underline{la}rge and \underline{s}mall \underline{co}llaboration framework for CSI feedback, referred to as \textbf{LASCO}.
LASCO operates by letting the base LAM produce an initial CSI reconstruction, learning the environment-induced reconstruction shift through a reference SAM and a proxy SAM, and transferring this shift back to the LAM.
To further enhance adaptability, we introduce elastic-LASCO (E-LASCO), which augments LASCO with learnable collaboration coefficients that control the contribution of LAMs and SAMs across different environments.
Numerical results demonstrate that LASCO and E-LASCO enables LAMs to achieve environment-specific performance gains with significantly reduced training costs, lower data collection requirements, and faster adaptation speed.

\end{abstract}

\section{Introduction}
Large artificial intelligence models (LAMs) have recently achieved remarkable breakthroughs across domains such as natural language processing and computer vision. Their capability to extract intricate patterns and learn high-dimensional representations provides a paradigm-shifting opportunity to address long-standing challenges in communication system design and optimization. Motivated by these advances, researchers have begun to explore the integration of LAMs into mobile communication systems, which is envisioned as a key enabler for 6G and beyond \cite{chen2024bigaimodels6g,yu2025aideeplearningterahertz}.

Recently, LAMs have been demonstrated with impressive performance for a wide range of tasks in wireless physical layers, including pre-trained LAM-based methods and native wireless LAM-based methods \cite{guo2025largeaimodelswireless}. Pre-trained LAM-based methods aim to leverage the generalized knowledge in pre-trained large language models (LLMs) or large vision models (LVMs) for physical layer tasks. With domain alignment between text/image data and wireless data, the pre-trained LAMs can be well adapted to different wireless tasks, including channel prediction \cite{liu2024llm4cp,li2025bridgingmodalitygapenhancing}, beam prediction \cite{sheng2025beam,zheng2025largelanguagemodelenabled,liu2025largemodelainearfield}, precoding \cite{zheng2025largelanguagemodelenabled}, CSI feedback \cite{cuileveraging}, port selection \cite{zhang2025portllmportpredictionmethod}, user association \cite{li2025jointuserassociationbeamforming}, equalization \cite{yu2025large}, reconfigurable intelligent surface (RIS) configurations \cite{huang2025llmrimsalargelanguagemodels} and more. Native wireless LAM-based methods typically incorporate numerous wireless data such as channel state information (CSI) to build an LAM from scratch, incentive the LAM to capture the intrinsic features of wireless channel. Some related works focus on generate a general representation of wireless channel for diverse downstream tasks \cite{alikhani2025largewirelessmodellwm,yang2025wirelessgpt}, while others dedicated to achieve high-performance and cross-environment generalization for a specific tasks \cite{guo2025promptenabledlargeaimodels,liu2025wifocfwirelessfoundationmodel}. 

\begin{figure*}[t]
    \centering
    \includegraphics[width=0.93\textwidth]{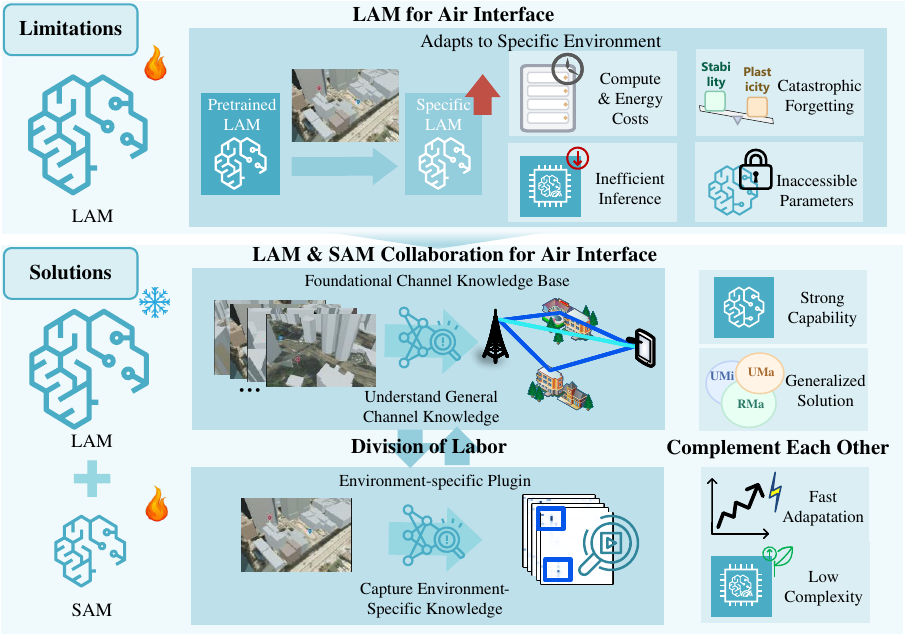}
        \caption{Illustration of the motivation of large and small model collaboration for air interface. A single LAM can hardly reap the performance gain brought by environment-specific adaptation. In the proposed framework, the large and small models are assigned complementary roles according to their respective characteristics, enabling a cooperative design that leverages their strengths while compensating for each other's limitations. LAM provides strong capability to offer generalized solutions across different environments, acting as a foundational channel knowledge base for understanding wireless channels. The SAM, with low complexity, can rapidly adapt to a target environment, acting as a plugin to capture environment-specific knowledge.}
	\label{fig:background}
\vspace{-0.5cm}
\end{figure*}

Generally, the aforementioned studies highlight the universal capability of LAMs to provide generalized solutions across diverse environments. However, most works overlook the potential performance gains of environment-specific adaptation, which remains crucial for practical deployment.
Despite the strong universality, LAMs still remarkably benefit from adaptation to downstream tasks \cite{liu2025structureawaredomainknowledgeinjection}, leading to better performance and a more cost-effective solution. Success has been achieved by fine-tuning the pre-trained LLMs with the datasets in specific field, such as medicine \cite{chen2023meditron70bscalingmedicalpretraining}, finance \cite{wu2023bloomberggptlargelanguagemodel}, and law \cite{huang2023lawyerllamatechnicalreport}. In wireless communications, the potential performance gain of adapting a model to specific environment has also been validated for small artificial intelligence models (SAMs) \cite{38843}. However, as illustrated in the upper part of Fig. \ref{fig:background}, environment adaptation of LAMs usually faces several critical challenges in practical deployment as follows.

First, owing to the extremely high computational complexity of training large models, environment-specific adaptation of LAMs often incurs substantial training delays and excessive energy consumption. This not only limits the practical feasibility of such adaptation but also leads to the aging of channel knowledge in dynamically varying environments, thereby diminishing the performance gains achieved through fine-tuning.
Second, since a base station (BS) typically serves dozens of UEs within each time slot, deploying separately fine-tuned LAMs for different user equipments (UEs) or regions results in very low inference efficiency due to hardware constraints and overwhelming memory consumption~\cite{DanielEtAl2023ContinuousBatchingLLM}.
Third, fine-tuning a large model on a specific domain can irreversibly overwrite its well-generalized knowledge, leading to catastrophic forgetting, particularly when the available dataset in the new environment is limited~\cite{Kirkpatrick2017EWC}.
Moreover, due to intellectual property restrictions, the parameters and architectures of LAMs are often inaccessible, rendering conventional fine-tuning-based adaptation infeasible.
Although recent studies have investigated low-rank adaptation (LoRA) techniques~\cite{hu2021loralowrankadaptationlarge}, these approaches still require full access to LAM parameters and remain inefficient for inference across multiple environments.

To overcome the aforementioned limitations, we advocate addressing the problems through a collaborative design between large and small models. As illustrated in the lower part of Fig.~\ref{fig:background}, the LAM, with its over-parameterized architecture, possesses strong representational power and can learn generalized channel knowledge from diverse datasets, but it is difficult to adapt efficiently to specific environments. In contrast, the SAM has lower complexity and can rapidly adapt to target environments with minimal overhead \cite{zhang2024fastslowgeneratingempirical}. Based on their respective characteristics, we designate the LAM as a static foundational channel knowledge base that provides generalized knowledge, and the SAM as a dynamic environment-specific plugin that captures environment-dependent knowledge. Through this division of labor, the two models complement each other, leveraging the LAM's generalization ability and the SAM's adaptability to achieve an effective synergy \cite{chen2025surveycollaborativemechanismslarge}. This collaborative framework is general and can be instantiated for a wide range of air interface tasks, such as channel prediction, beam management, channel estimation, and more, enabling different tasks to benefit from environment-specific performance gains.

Building on this collaborative philosophy, we instantiate it in the context of CSI feedback by proposing LASCO, a \underline{\textbf{la}}rge and \underline{\textbf{s}}mall \underline{\textbf{co}}llaboration framework for CSI feedback.
LASCO concretizes the division-of-labor principle introduced above. While assigning the LAM as a static knowledge base that explores universal channel knowledge, the key idea is to emulate the effect of environment adaptation, that is, the adaptation-induced output distribution shift of the LAM, through lightweight SAMs. To achieve this, LASCO employs two functionally complementary SAMs: a \emph{reference SAM}, aligned with the base LAM during pre-training to mimic the pretrained behavior, and a \emph{proxy SAM}, fine-tuned on target-environment data to emulate the adapted behavior. The discrepancy between their outputs represents the estimated distribution shift and is applied to the LAM to realize adaptation. Using two SAMs rather than one preserves task-level consistency with the LAM: both networks remain focused on CSI reconstruction, allowing the proxy SAM to inherit and reuse the pretrained representations instead of relearning from scratch. 
Furthermore, we observe that the optimal collaboration pattern between large and small models varies across different environments. To accommodate such diversity, we extend the proposed LASCO framework to an elastic form, referred to as elastic-LASCO (E-LASCO), which introduces learnable hyper-parameters to adaptively adjust the collaboration strength between the large and small models in each environment. This elastic design enables environment-specific cooperation and further improves the overall CSI feedback performance.
Numerical results verify that LASCO attains the performance gain of environment adaptation with low complexity, no parameter access, and accelerated convergence.

This paper makes the following key contributions:
\begin{itemize}
    \item We analyze the complementary roles of large and small models in air interface intelligence and establish a general collaboration framework. In this framework, LAMs act as foundational channel knowledge bases that provide a generalized understanding of channel characteristics, whereas SAMs serve as lightweight, environment-specific plugins that enable rapid adaptation.    
    
    \item We develop LASCO, a large-small collaboration framework that instantiates the proposed philosophy in the context of CSI feedback. LASCO realizes the division-of-labor principle by designating the LAM as a static foundational knowledge base and introducing two complementary SAMs to emulate the adaptation-induced output distribution shift of the LAM across environments, thereby enabling computationally efficient, environment-specific adaptation for LAM-based CSI feedback.

    \item We further extend LASCO to an elastic form, termed E-LASCO, to accommodate diverse collaboration behaviors across different environments. E-LASCO introduces learnable hyperparameters that adaptively balance the contributions of large and small models, thereby enhancing environment-specific performance.

    \item Extensive simulations verify the effectiveness of the proposed frameworks. Both LASCO and E-LASCO consistently outperform baseline methods, achieving more accurate CSI reconstruction, improved adaptation capability, and significantly reduced data collection and training overhead.
\end{itemize}

The remainder of this paper is organized as follows. Section II analyzes the necessity of collaboration between LAMs and SAMs in the air interface and presents a general large-small model collaboration framework. Section III develops LASCO and its elastic extension E-LASCO, which instantiate the proposed framework for CSI feedback. Section IV provides extensive numerical results to demonstrate the effectiveness of the proposed frameworks. Finally, Section V concludes the paper.

\section{Large and Small Model Collaboration}
\label{sec:framework}
This section presents the theoretical foundation and general design of the proposed LAM-SAM collaboration framework. We first analyze the complementary properties of large and small models and then introduce the unified architecture that integrates generalized and environment-specific knowledge.

\subsection{Motivation and Conceptual Foundation}
\label{sec:motivation}
AI-based air interface design has attracted considerable interest in both academia and standardization communities \cite{213599,234039,251870,251881}. Existing approaches typically rely on either LAMs or SAMs. While each type achieves promising performance, both exhibit intrinsic limitations when deployed individually. We explain why LAMs and SAMs should collaborate and analyzes their complementary properties.

\textbf{1) Why LAMs need SAMs?} LAMs usually feature with a large set of neural parameters, which limits its training efficiency and flexibility. We highlight the two key advantages of introducing SAMs as collaborators to LAMs. First, the performance-complexity tradeoff is a critical issue for AI-based air interfaces. Fine-tuning an LAM for each target environment involves updating millions of parameters, resulting in slow adaptation and energy-intensive training. The learned channel knowledge may also become outdated before completion, degrading the performance gain brought by environment-specific adaptation. The reason behind is that LAMs are usually inherently over-parameterized, which encode wireless environments with highly redundant representations to achieve a generalized solution \cite{nakkiran2019deepdoubledescentbigger}. However, environment-specific adaptation often lies in a much lower-dimensional subspace. According to the concept of intrinsic dimension \cite{aghajanyan-etal-2021-intrinsic}, the effective degrees of freedom required for environment adaptation might be far smaller than the parameter count of an LAM. Using the full LAM to capture this low-dimensional shift is computationally wasteful, whereas an SAM can efficiently learn the essential variations within a compact parameter space. The SAM thus provides a cost-effective path to rapid environment adaptation without sacrificing performance.

Second, SAMs bring better stability-plasticity tradeoff to LAMs. LAMs are usually pre-trained to capture generalized channel knowledge. Fine-tuning the LAMs with the datasets collected in specific environments might cause catastrophic forgetting \cite{Parisi2019CLReview} of the generalized channel knowledge achieved by pre-training, especially when the environment-specific datasets are deficient. By introducing external SAMs to capture the environment-specific knowledge, the global knowledge retention and local adaptation can be decoupled, which helps the LAM preserve stable global representations. This cooperative division of labor mitigates forgetting and ensures both stability and plasticity during continuous deployment.

\textbf{2) Why SAMs need LAMs?} Due to the limited model scales, SAMs usually fall into under-parameterized or critically parameterized regime in wireless communication tasks, which may cause wireless hallucination and unfavorable generalization when dealing with complicated communication scenarios in deployment. Introducing LAMs as collaborators to SAMs bring the following advantages. First, LAMs can help reduce wireless hallucination in SAMs. Shallow SAMs tend to approximate objectives through oversimplified mappings. While achieving favorable performance in specific environments, the SAMs often producing outputs that violate physical propagation constraints and cause wireless hallucination \cite{wang2025wirelesshallucinationgenerativeaienabled}. The LAM, trained on massive and heterogeneous datasets, encodes the underlying physical regularities of the channel and provides physically consistent priors for downstream tasks. With this guidance, the LAMs and SAMs can collaborate to generate outputs that simultaneously manifests the environment specific features and better obeys the physical rules.

Second, SAMs usually lack the capacity to generalize across diverse environments. Their limited parameterization constrains the loss landscape to only a few sharp or highly specialized minima, making it difficult for stochastic optimizers such as stocastic gradient descent or AdamW to reach flat and stable solutions that typically yield good generalization \cite{Hochreiter1997FlatMinima,Keskar2017SharpMinima}. Consequently, when dealing with heterogeneous channel conditions, SAMs tend to underfit or converge to brittle minima, resulting in unstable cross-scenario performance. In contrast, the LAM operates in a highly over-parameterized regime, where large-scale pre-training and stochastic optimization naturally guide the model toward wide low-curvature basins that encode universal channel patterns shared across environments \cite{Kaplan2020Scaling}. These implicitly regularized and highly general features form a universal channel knowledge base. By collaborating with the LAM, SAMs can leverage this cross-environment robustness while focusing their limited capacity on efficient environment-specific refinement.

\begin{figure*}[t]
    \centering
    \includegraphics[width=0.96\linewidth]{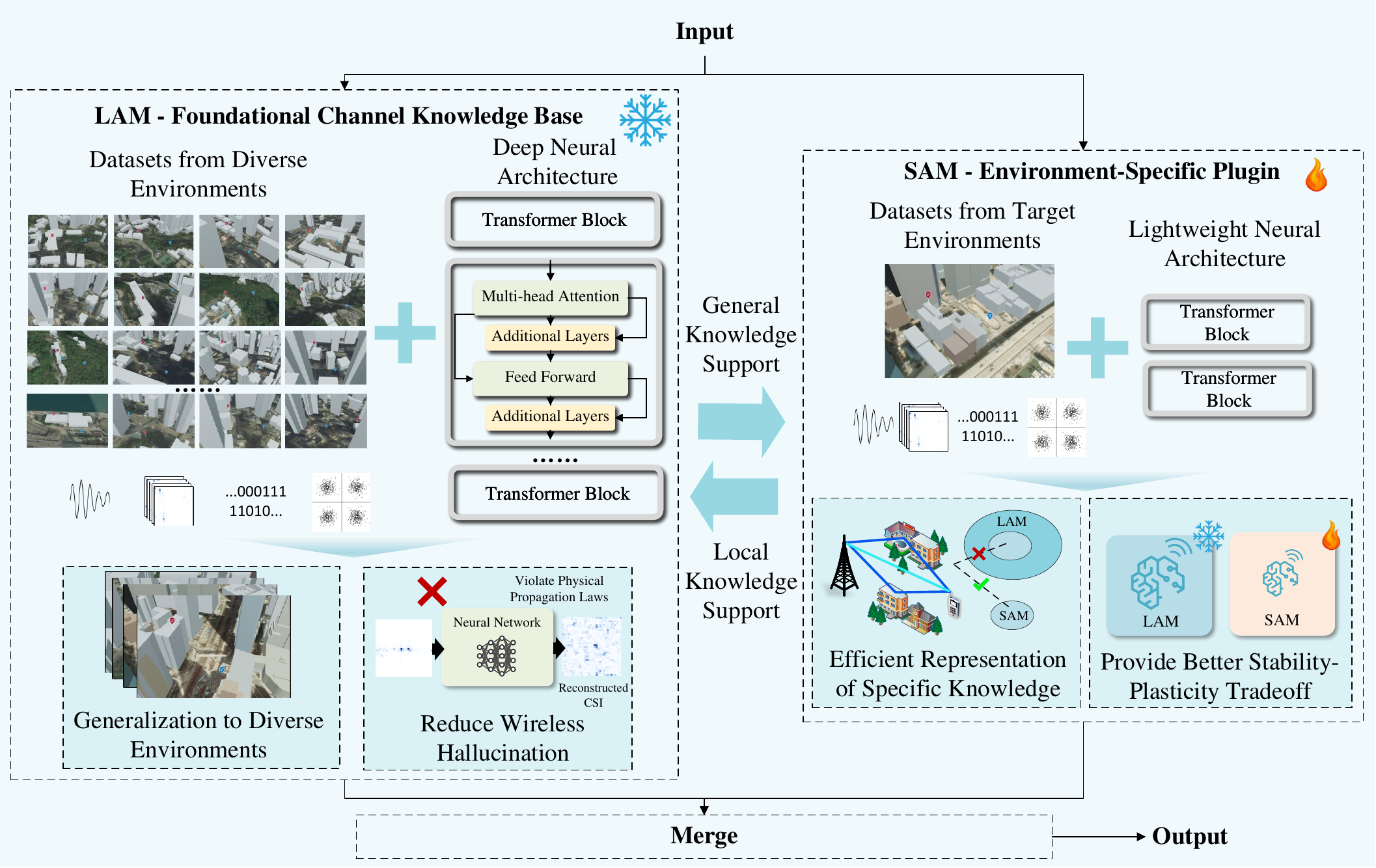}
    \caption{Illustration of the proposed large-small model collaboration framework for the air interface. 
    The LAM serves as a foundational channel knowledge base, pre-trained on heterogeneous datasets using a deep Transformer architecture to capture generalized propagation knowledge and provide global priors. 
    The SAM acts as an environment-specific plugin, trained on target-environment data with a lightweight neural architecture to refine the LAM's representation. 
    Through bidirectional knowledge flow, the LAM supplies generalized priors while the SAM contributes local adaptations. 
    Their outputs are fused into environment-adaptive channel representations, improving physical consistency, environment fidelity, and the overall stability-plasticity balance.}
    \label{fig:general_framemwork}
\end{figure*}

\subsection{General Framework Design}
\label{sec:framework_design}

Building on the above analysis, we then introduce the overall design of the proposed large and small model collaboration framework. The framework is built on the idea that general channel knowledge and local environment-specific knowledge can be handled by different components. Instead of letting a single model learn both kinds of knowledge, we separate the extraction of universal wireless information and the representation of environment-dependent characteristics. This design philosophy is reflected in Fig. \ref{fig:general_framemwork}, which visualizes the knowledge flow between the LAM-based foundational channel knowledge base and the SAM-based environment-specific plugin.

Consider an air interface task that maps an input $\mathbf{p}$ to an output $\mathbf{q}$, where $\mathbf{p}$ may represent pilot symbols, received CSI measurements, or other task-related features, and $\mathbf{q}$ denotes the desired output such as reconstructed CSI, predicted channel coefficients, or detected symbols. A dataset collected in environment $\mathcal{S}$ is denoted by
\begin{equation}
    \mathcal{D}_{\mathcal{S}}
    = \{ (\mathbf{p}_i,\mathbf{q}_i) \}_{i=1}^{N_{\mathcal{S}}}.
\end{equation}
where $N_{\mathcal{S}}$ represents the number of samples in the dataset $\mathcal{D}_{\mathcal{S}}$.

To provide a stable basis for air interface tasks, the framework first constructs a foundational channel knowledge base using LAMs, which is illustrated in the left part of Fig. \ref{fig:general_framemwork}. This component is responsible for capturing broad spatial-frequency patterns and robust propagation structures that remain consistent across diverse scenarios. A deep Transformer is adopted as the backbone, which scales favorably with model size and becomes the de facto backbone for most large foundation models. Also, its global receptive field captures dependencies across antennas, subcarriers, and multi-path components. To ensure that the LAMs learn environment-agnostic knowledge rather than memorizing local features, the LAMs are pre-trained on a large mixed dataset collected from multiple environments. After pre-training, the LAM produces a generalized representation or preliminary solutions. The foundational channel knowledge base can be formulated as follows:
\begin{equation}
    \tilde{\mathbf{q}}_{\mathrm{kb}} = f_{\mathrm{LAM}}(\mathbf{p}),
\end{equation}
where $f_{\mathrm{LAM}}(\cdot)$ denotes the LAM and $\tilde{\mathbf{q}}_{\mathrm{kb}}$ represents the output of the foundational channel knowledge base.

Although this foundational channel knowledge base provides generalized knowledge, there are also local propagation characteristics that differ across environments. To capture these variations, the framework incorporates an environment-specific plugin based on lightweight SAMs. As shown in the right part of Fig. \ref{fig:general_framemwork}, the SAM processes the same input and learns to adjust the LAM's generalized output according to data collected in the target environment. Because environment-specific variations are typically lower in complexity, the SAM adopts a compact neural architecture formulated as follows:
\begin{equation}
    \tilde{\mathbf{q}}_{\mathrm{sp}} = f_{\mathrm{SAM}}(\mathbf{p}),
\end{equation}
where $f_{\mathrm{SAM}}(\cdot)$ is a small model that will be adapted when the environment changes, and $\tilde{\mathbf{q}}_{\mathrm{sp}}$ represents the output of the environment-specific plugin. The roles of the two models are distinct. The LAM encodes stable knowledge that should remain unchanged across environments, and the SAM learns the component that must adjust across deployments. This separation allows the SAM to respond quickly to environmental changes without disturbing the generalized channel representations already learned by the LAM.

To obtain the final task output, the framework integrates the generalized solution from the LAM and the environment-specific solution from the SAM through a merging function. 
This function performs knowledge fusion and can be implemented either as simple arithmetic operations, such as addition or weighted averaging, or as a small learnable neural module depending on the task design. The merging function generates the final prediction as follows:
\begin{equation}
    \mathbf{q} = g\!\left(\tilde{\mathbf{q}}_{\mathrm{kb}},\tilde{\mathbf{q}}_{\mathrm{sp}}\right),
\end{equation}
where $g(\cdot)$ denotes the merging function. Through this process, global propagation knowledge and local environmental cues are combined into a unified and adaptive output.

The adaptation workflow aligns naturally with this structure. 
The foundational LAM is trained offline using heterogeneous datasets and its parameters remain frozen in deployment to preserve the generalized channel knowledge it encodes. 
When the environment changes, new data are collected in the target environment to form $\mathcal{D}_\mathcal{S}$, and the SAMs in the environment-specific plugin are fine-tuned. The adaptation objective is given by
\begin{equation}
    f_{\mathrm{SAM}}^\star
    =
    \argmin_{f_{\mathrm{SAM}}}
    \mathbb{E}_{(\mathbf{p},\mathbf{q})\in\mathcal{D}_\mathcal{S}}
    \Big\{
        l\!\left(
        \mathbf{q},
        g\!\left(
            f_{\mathrm{LAM}}(\mathbf{p}),
            f_{\mathrm{SAM}}(\mathbf{p})
        \right)
        \right)
    \Big\},
\end{equation}
where $l(\cdot)$ is a task-specific loss such as MSE for reconstruction or cross entropy for detection. The proposed LAM-SAM collaboration is a task-agnostic framework that can be extended to a wide range of tasks in air interface, including channel prediction, channel estimation, beam management, CSI feedback, and more. For each task, suitable choices of the LAM architecture, SAM architecture, and merging function allow the framework to reap the advantages of efficient environment-specific adaptation, thus effectively pushing the performance frontier of AI air interface design.

\subsection{Discussion}
Despite the aforementioned advantages, we further discuss the computational complexity and intellectual property protection of the proposed large and small model collaboration framework.

In environment-specific adaptation, the computational cost is dominated by backpropagation through network parameters. In conventional LAM-based solutions, fine-tuning requires updating large numbers of parameters, resulting in high memory usage, long training latency, and significant energy consumption. In contrast, our framework freezes the LAM entirely and fine-tunes only the lightweight SAM, whose parameter count is orders of magnitude smaller. Since the LAM and the SAM operate in a parallel way, this design eliminates the need for backpropagation through deep neural architecture and reduces the training complexity to that of a small model. Consequently, adaptation can be completed with limited computational resources, enabling fast and frequent environment updates.

During inference, LAM-based approaches that customize a separate LAM for each UE or each environment incur extremely high computational and memory overhead. When serving dozens of UEs simultaneously, due to the difference in the LAM parameters, the inference cannot be combined into a large batch for efficient calculation. Instead, multiple LAMs need to operate in a small batch or even batch-1 mode, causing poor hardware utilization and rapid memory exhaustion when the number of UEs increases. Under the proposed framework, all users share the same pre-trained LAM, which can be executed once in a batched manner. Only the lightweight SAMs, one per environment or per UE group, are executed individually. This significantly reduces graphics processing unit (GPU) memory consumption and increases computation efficiency. The advantage becomes more prominent as the number of simultaneously served UEs grows, which is aligned with the deployment characteristics of massive MIMO systems.

Finally, a crucial benefit of the proposed framework is that collaboration between LAMs and SAMs does not require access to the internal parameters and architecture. Only the outputs of the LAM are needed for SAM training and for performing distribution-shift imitation. This makes the framework compatible with commercial or proprietary LAMs whose weights cannot be exposed due to intellectual property, licensing, or security restrictions. From a practical perspective, this property greatly simplifies the deployment of LAM-based air interface intelligence in multi-vendor systems and enable the use of close-source off-the-shelf wireless LAMs for environment-specific adaptation. This black-box nature of LAM access can also simplify the standardization of LAM-SAM collaboration \cite{38843}.

\section{Case Study: A Large and Small Model Collaboration for CSI Feedback}
\label{sec:lasco}

To demonstrate the practical value of the proposed framework, we apply it to the CSI feedback task. The resulting design, termed LASCO, instantiates the LAM-SAM collaboration concept in a concrete wireless context.

\subsection{Channel and Signal Model}

We consider a downlink massive antenna array scenario in which the BS is equipped with a uniform linear array (ULA) of \(N_{\mathrm{t}}\) antenna elements (\(N_{\mathrm{t}}\gg1\)) and serves a single-antenna UE. Transmission is based on an orthogonal frequency division multiplexing (OFDM) scheme with \(N_{\mathrm{c}}\) subcarriers. The channel on the \(i\)-th subcarrier is modeled using a clustered multipath representation:
\begin{equation}  
\widetilde{\mathbf{h}}_i=\sum_{p=1}^{N_{\mathrm{p}}}\sum_{s=1}^{N_{\mathrm{s}}}\alpha_{i,p,s}\,\mathbf{a}(\theta_{i,p,s}),
\end{equation}
where \(N_{\mathrm{p}}\) denotes the number of scattering clusters, each containing \(N_{\mathrm{s}}\) subpaths. The complex path gain for the \(s\)-th subpath of cluster \(p\) is \(\alpha_{i,p,s}\), and \(\theta_{i,p,s}\) is its angle of departure (AoD). For a ULA the steering vector is written as
\begin{equation}
    \mathbf{a}(\theta)=\big[1,\ e^{j2\pi\frac{\Delta}{\lambda}\sin\theta},\ \ldots,\ e^{j2\pi\frac{(N_{\mathrm{t}}-1)\Delta}{\lambda}\sin\theta}\big]^\mathsf{T},
\end{equation}
where \(\Delta\) is the antenna spacing and \(\lambda\) the carrier wavelength. The received scalar on subcarrier \(i\) is given by the standard narrowband downlink model:
\begin{equation}
    y_i=\mathbf{h}_i^H\mathbf{v}_i x_i + z_i,
\end{equation}
with \(\mathbf{h}_i\in\mathbb{C}^{N_{\mathrm{t}}\times 1}\) the downlink channel for subcarrier \(i\), \(\mathbf{v}_i\in\mathbb{C}^{N_{\mathrm{t}}\times 1}\) the beamformer, \(x_i\in\mathbb{C}\) the transmitted symbol and \(z_i\) complex AWGN. Collecting all subcarrier channels produces the full-frequency CSI matrix
\[
\mathbf{H}=[\mathbf{h}_1,\mathbf{h}_2,\ldots,\mathbf{h}_{N_{\mathrm{c}}}]\in\mathbb{C}^{N_{\mathrm{t}}\times N_{\mathrm{c}}},
\]
which corresponds to \(2N_{\mathrm{t}}N_{\mathrm{c}}\) real-valued parameters when the complex entries are represented by their real and imaginary parts.

\subsection{DL-based CSI Feedback Pipeline}
To reduce the uplink feedback burden, we adopt a deep learning-based CSI feedback scheme. To alleviate the potential problems caused by the multi-vendor collaboration for two-side models, we focus on one-side model in this paper. In this scheme, after the UE obtains \(\mathbf{H}\) (converted to a real-valued vector), a random linear projection is applied to produce a low-dimensional codeword \(\mathbf{s}\in\mathbb{R}^{M\times 1}\):
\begin{equation}
    \mathbf{s} = \mathbf{A}\,\mathrm{vec}(\mathbf{H}),
\end{equation}
where \(\mathbf{A}\in\mathbb{R}^{N_{\mathrm{s}}\times 2N_{\mathrm{t}}N_{\mathrm{c}}}\) is a random projection matrix and \(\mathrm{vec}(\cdot)\) denotes vectorization of the real-valued CSI representation. The compression ratio is therefore defined as
\begin{equation}
    \gamma=\frac{2N_{\mathrm{t}}N_{\mathrm{c}}}{M}.
\end{equation}
Upon reception of \(\mathbf{s}\) at the BS, a coarse estimate of the original CSI is obtained by applying the Moore-Penrose pseudo-inverse of \(\mathbf{A}\) and reshaping back into matrix form:
\begin{equation}
    \mathbf{H}_{\mathrm{in}}=\mathrm{devec}(\mathbf{A}^{\dagger}\mathbf{s}).
\end{equation}
While the initial reconstructed CSI might not accurate using only linear processing, this initial reconstruction is then refined by a neural network:
\begin{equation}
    \widehat{\mathbf{H}} = f_{\mathrm{NN}}(\mathbf{H}_{\mathrm{in}}),
\end{equation}
where \(f_{\mathrm{NN}}(\cdot)\) denotes the learned enhancement mapping and \(\widehat{\mathbf{H}}\) is the final recovered CSI. 

To quantify reconstruction performance, we employ commonly used loss measures. The MSE is used to evaluate reconstruction fidelity:
\begin{equation}
    l_{\mathrm{MSE}}(\mathbf{H},\widehat{\mathbf{H}})=\frac{\|\widehat{\mathbf{H}}-\mathbf{H}\|_2^2}{\|\mathbf{H}\|_2^2},
    \label{equ:mse}
\end{equation}
which captures the relative power of the reconstruction error. Because accurate CSI is primarily needed for beamforming, we also consider the negative generalized cosine similarity (GCS) metric that measures alignment between the true and reconstructed per-subcarrier channel vectors:
\begin{equation}
    l_{\mathrm{GCS}}(\mathbf{H},\widehat{\mathbf{H}}) 
    = -\frac{1}{N_{\mathrm{c}}}\sum_{i=1}^{N_{\mathrm{c}}}\frac{|\widehat{\mathbf{h}}_i^H\mathbf{h}_i|}{\|\widehat{\mathbf{h}}_i\|_2\|\mathbf{h}_i\|_2},
    \label{equ:cs}
\end{equation}
where \(\widehat{\mathbf{h}}_i\) denotes the reconstructed channel for subcarrier \(i\).

\begin{figure*}[t]
	\centering
    \includegraphics[width=0.93\linewidth]{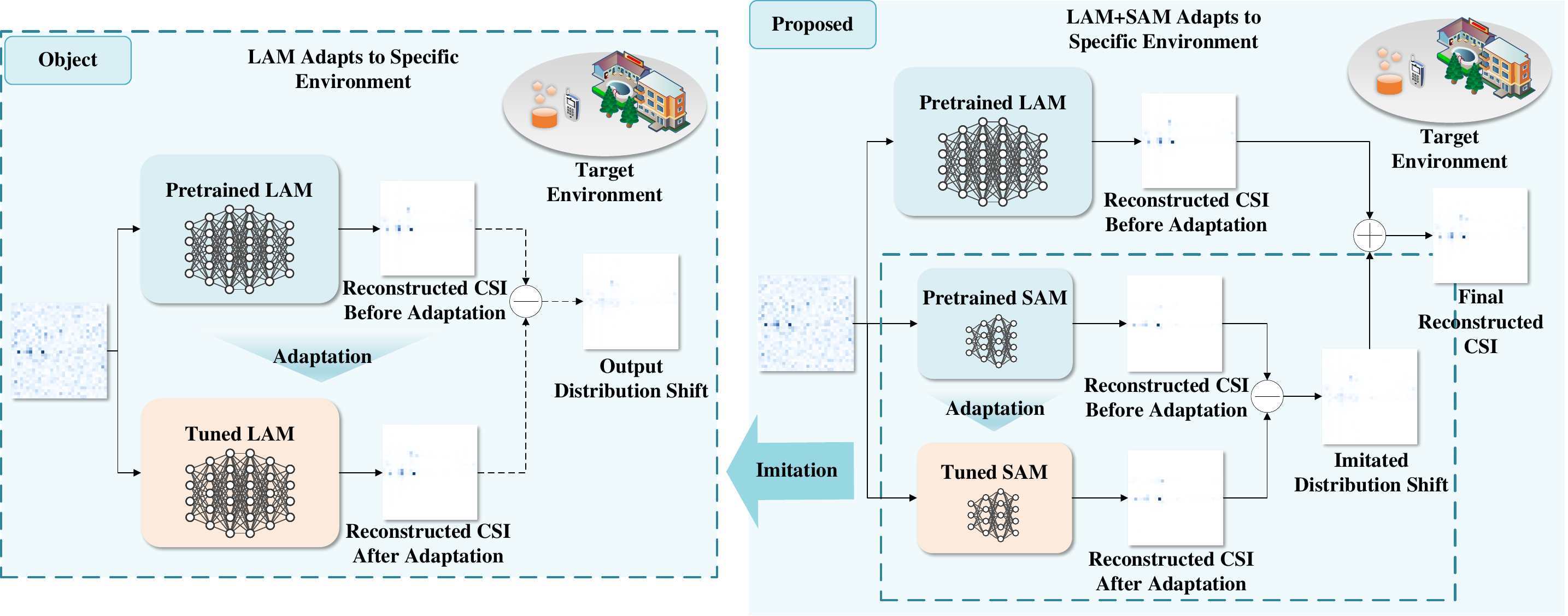}
	\caption{Illustration of the motivation of LASCO. The object is to imitate the LAM's environment adaptation behavior. In the proposed framework, we attempt to approximate the LAM's output distribution shift caused by environment adaptation with two SAMs, and exerted the imitated distribution shift to the LAM to emulate the impact of environment adaptation.}
	\label{fig:mimic}
	\vspace{-0.6cm}
\end{figure*}

\subsection{Key Framework}
Efficient DL-based CSI feedback is usually dependent on a good understanding of the propagation environment. The performance of CSI feedback can usually significantly benefit from the environment-specific adaptation. In traditional environment adaptation pipelines for CSI feedback tasks, an LAM must be fine-tuned on CSI dataset from the target environment to achieve environment-specific performance \cite{zeng2021downlink}. While effective in principle, this approach faces significant deployment barriers: fine-tuning an LAM for CSI feedback incurs substantial computational complexity and requires direct access to its parameters, both of which limit its practicality. As a result, assigning the adaptation task directly to the LAM itself is neither efficient nor deployment-friendly. Moreover, because the LAM is inherently over-parameterized, it learns rich and generalized representations of wireless environments that capture underlying physical and statistical structures. These generalized representations constitute valuable knowledge that should be preserved and reused, rather than modified through fine-tuning. From this perspective, we advocate treating the LAM as a fixed foundational channel knowledge base, serving as a stable repository of generalized channel understanding, and focusing on exploiting its knowledge through external mechanisms rather than internal retraining.

To achieve environment adaptation without modifying the LAM, we adopt a new viewpoint: rather than updating the model parameters, we approximate the result of adaptation. As illustrated in the left part of Fig. \ref{fig:mimic}, fine-tuning leads to a change in the LAM's output distribution, that is, the distribution of reconstructed CSI, making it better aligned with the target environment. Therefore, if the LAM's parameters are inaccessible, one can instead approximate this output distribution shift externally and impose it on the LAM's outputs, thereby mimicking the behavior of fine-tuning \cite{liu2024tuninglanguagemodelsproxy,he2024cptconsistentproxytuning}. This output distribution shift embodies the environment-specific knowledge that distinguishes one scenario from another. Unlike the complex and generalized knowledge encoded by the LAM, such environment-specific knowledge is relatively simple and localized, making it more amenable to modeling by lightweight neural networks. Accordingly, we introduce SAMs as a dynamic plugin that captures this environment-dependent knowledge efficiently and flexibly.

As depicted in the right part of Fig. \ref{fig:mimic}, to mimic the adaptation-induced output distribution shift of the LAM, we need to: 1) approximate the output distribution of the pre-trained LAM, and 2) approximate its post-adaptation output distribution.
We denote the pre-trained, multi-environment model as the base LAM, and introduce a compact reference SAM to emulate its output behavior. The reference SAM employs a similar backbone architecture and is trained on the same mixed datasets using comparable training strategies, ensuring that its output distribution closely aligns with that of the base LAM. To represent the LAM's adaptation behavior, we introduce another model of identical structure, termed the proxy SAM. The proxy SAM is initialized with the parameters of the reference SAM, providing a comparable starting point to the pre-trained LAM. The proxy SAM is then fine-tuned on CSI dataset from the target environment to emulate the adapted output distribution. The difference between the proxy SAM and the reference SAM thus captures an approximation of the LAM's output distribution shift after adaptation. By applying this estimated shift to the outputs of the base LAM, we effectively reproduce the adaptation effect of fine-tuning, without accessing or retraining the LAM itself. 

Using two small models rather than one is essential to preserve task-level consistency and facilitate knowledge transfer. If a single SAM were trained only to correct the residual between the ground-truth CSI and the LAM's reconstructed CSI, its learning objective would deviate from the original CSI reconstruction task, preventing effective reuse of the pre-trained representations. By maintaining a reference SAM that mirrors the pre-trained behavior of the LAM, and a proxy SAM that fine-tunes toward the adapted behavior, the proposed dual-SAM scheme allows the adaptation process to stay aligned with the original reconstruction task while capturing environment-specific variations efficiently. Formally, LASCO can be formulated as follows:
\begin{equation}
    \widehat{\mathbf{H}}=f_\mathrm{base}(\mathbf{H}_\mathrm{in})+\underbrace{f_\mathrm{pxy}(\mathbf{H}_\mathrm{in})-f_\mathrm{ref}(\mathbf{H}_\mathrm{in})}_{\text{env. specific variation}}.
    \label{equ:lsc}
\end{equation}
where $f_\mathrm{base}(\cdot)$, $f_\mathrm{pxy}(\cdot)$, and $f_\mathrm{ref}(\cdot)$ represents the base LAM, the proxy SAM, and the reference SAM, respectively.

However, although the reference SAM is well aligned with LAM, there are inevitable differences between the SAM and the LAM due to the heterogeneity in neural architectures. To alleviate the performance loss caused by the difference, we further consider the participation of LAM in the environment adaptation process. Specifically, compared to directly fine-tuning the proxy SAM, we consider the co-inference of the three models in fine-tuning process \cite{liu2024tuninglanguagemodelsproxy}, and modify the loss function \ref{equ:mse} of the proxy SAM fine-tuning as follows:
\begin{equation} 
    l_\mathrm{1}=\|\mathbf{H}-(f_\mathrm{base}(\mathbf{H}_\mathrm{in})+f_\mathrm{pxy}(\mathbf{H}_\mathrm{in})-f_\mathrm{ref}(\mathbf{H}_\mathrm{in}))\|_2^2.
    \label{equ:loss_init}
\end{equation}

\begin{figure}[t]
	\centering
	\subfigure[Pre-training]{\includegraphics[width=0.93\linewidth]{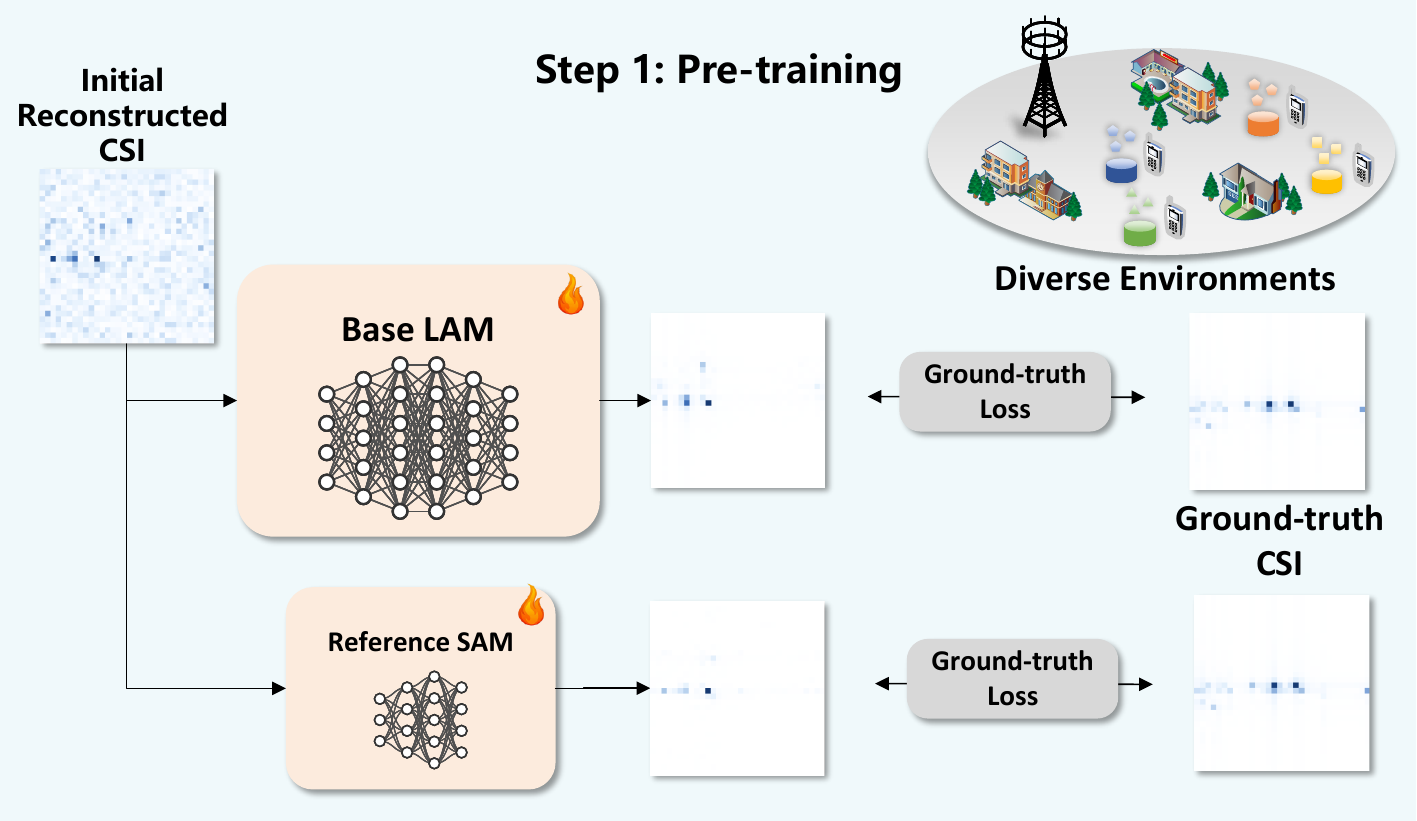}}
	\subfigure[Environment adaptation]{\includegraphics[width=0.93\linewidth]{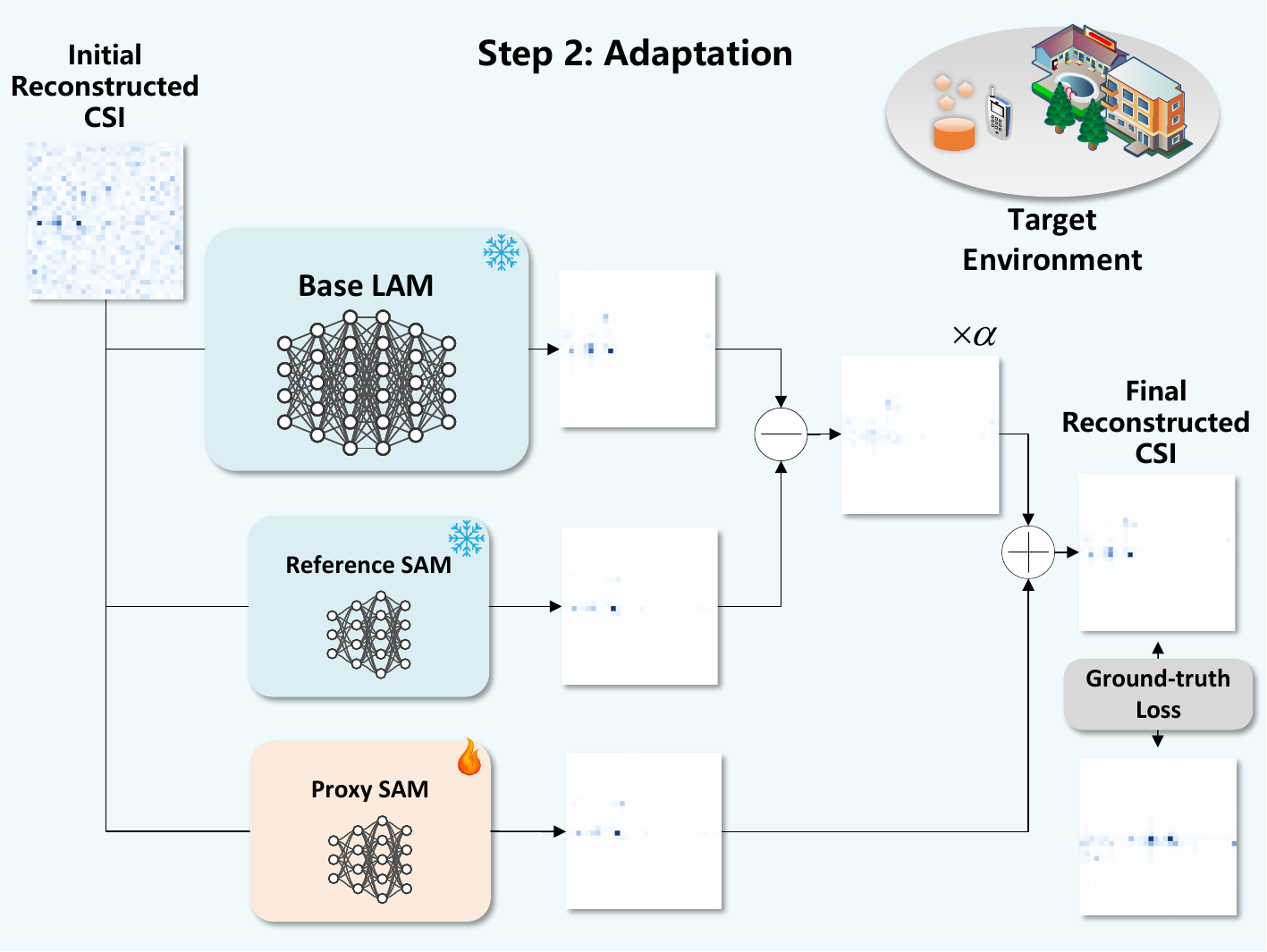}} 
	\caption{Illustration of the detailed training steps of LASCO. (a) \textbf{Pre-training}: A base LAM is first built with datasets collected in diverse environments. A reference SAM is aligned with the base LAM with using identical training strategies with the base LAM. (b) \textbf{Environment Adaptation}: When adapting to new environments, a proxy SAM initialized with the parameters of reference SAM is fine-tuned with the awareness of the inference time collaboration.}
	\label{fig:steps}
	\vspace{-0.6cm}
\end{figure}

In practical deployment, the collaboration between the LAM and SAMs is also impacted by the characteristics of different environments. The CSI in certain environments might be accurately reconstructed by the LAM and less rely on environment-specific adaptation, while other CSI might deviate from common distribution and more benefit from fine-tuning. Therefore, we introduce a hyper-parameter $\alpha$ to better control the collaboration between the LAM and the SAM. The loss function can be reformulated as follows:
\begin{equation} 
    l_\mathrm{2}=\|\mathbf{H}-(f_\mathrm{base}(\mathbf{H}_\mathrm{in})+\alpha(f_\mathrm{pxy}(\mathbf{H}_\mathrm{in})-f_\mathrm{ref}(\mathbf{H}_\mathrm{in})))\|_2^2.
    \label{equ:loss_modify}
\end{equation}
Because the proxy SAM is a trainable network, directly multiplying its output by a scaling factor $\alpha$ may fail to achieve the intended modulation effect, as the model can implicitly compensate for the scaling during training. To address this issue, we refine (\ref{equ:lsc}) to incorporate a more effective adjustment mechanism that preserves training stability while maintaining control over the adaptation strength as follows:
\begin{equation}
    \widehat{\mathbf{H}}=f_\mathrm{pxy}(\mathbf{H}_\mathrm{in})+\alpha(f_\mathrm{base}(\mathbf{H}_\mathrm{in})-f_\mathrm{ref}(\mathbf{H}_\mathrm{in})),
\end{equation}
and the final optimization object is correspondingly modified as follows:
\begin{equation}
\min_{f_\mathrm{pxy}}
\left\| 
\mathbf{H} -
\big( 
f_\mathrm{pxy}(\mathbf{H}_\mathrm{in}) + 
\alpha \big( f_\mathrm{base}(\mathbf{H}_\mathrm{in}) - f_\mathrm{ref}(\mathbf{H}_\mathrm{in}) \big)
\big)
\right\|_2^2 .
\label{eq:loss_modified2}
\end{equation}
Specifically, as shown in Fig. \ref{fig:steps}, LASCO includes two training steps as follows:
\begin{itemize}
   \item \textbf{Step 1-Pre-training:} The base LAM and the reference SAM are pre-trained with the same mixed CSI datasets collected from diverse environments. After being pre-trained with diverse CSI datasets, the base LAM can learn the general channel knowledge and act as a general channel knowledge base, and the SAM is well aligned with the LAM.
   \item \textbf{Step 2-Environment Adaptation:} The proxy SAM is first initialized with the parameters of the reference SAM, and then fine-tuned with the CSI datasets in the target environment according to (\ref{equ:loss_modify}). The parameters of the base LAM and the reference SAM are kept frozen during fine-tuning.
\end{itemize}

\subsection{Elastic Large and Small Model Collaboration for CSI Feedback}

Although the collaboration between the LAM and SAMs enables efficient environment adaptation, the optimal strength of their cooperation is not universal across different environments. In some cases, the pre-trained LAM already provides a sufficiently accurate reconstruction, and excessive reliance on the SAM may introduce unnecessary modifications. In other cases, the target environment diverges considerably from the conditions covered during pre-training, and the SAM must contribute more prominently to compensate for the mismatch. Therefore, a fixed collaboration coefficient $\alpha$ cannot appropriately capture such diverse dependencies between generalized and environment-specific knowledge.

To overcome this limitation, we extend LASCO into an elastic large and small model collaboration framework, referred to as E-LASCO. In this framework, the cooperation strength between the LAM and the SAM is not manually determined but learned automatically during the adaptation process. Instead of assigning $\alpha$ as a predefined constant, we treat it as an optimization variable jointly trained with the parameters of the proxy SAM. Formally, the environment adaptation problem in E-LASCO can be expressed as
\begin{equation}
\min_{f_\mathrm{pxy},\, \alpha}
\left\| 
\mathbf{H} -
\big( 
f_\mathrm{pxy}(\mathbf{H}_\mathrm{in}) + 
\alpha \big( f_\mathrm{base}(\mathbf{H}_\mathrm{in}) - f_\mathrm{ref}(\mathbf{H}_\mathrm{in}) \big)
\big)
\right\|_2^2 .
\label{eq:elasco_opt}
\end{equation}
By introducing $\alpha$ into the optimization process, E-LASCO allows the model to automatically adjust the contribution of generalized knowledge from the LAM and environment-specific refinements from the SAM, achieving an adaptive balance suited to the current environment.

This design has clear practical benefits for deployment. In LASCO, $\alpha$ must be selected manually, which requires costly hyperparameter search in each new environment. Since the optimal $\alpha$ varies considerably across scenarios, an inappropriate choice may cause substantial performance degradation, which will be further demonstrated in the following section according to numerical results. In practical deployments, this search process is infeasible, as the true environment statistics are often unavailable beforehand. By contrast, E-LASCO eliminates the need for manual tuning: the collaboration weight $\alpha$ is learned jointly with the adaptation network, enabling the model to autonomously infer the optimal collaboration pattern based on observed data. In this sense, the learnable $\alpha$ enhances not only the adaptability but also the usability of the overall framework, reducing the engineering burden and improving robustness under uncertain deployment conditions.

\subsection{Neural Architecture}
For both LAMs and SAMs, Transformers are adopted as basic neural architecture \cite{vaswani2017attention}. The base LAM is designed as a foundational channel knowledge base for CSI feedback tasks. Therefore, a deep Transformer network with 20 Transformer blocks is employed for the base LAM. The $d_\mathrm{model}$ and the dimension of the feed-forward module are set to 512 and 2048, respectively. Pre-normalization is used for LAM, which benefits from its optimization stability.
The reference SAM and the proxy SAM are designed to capture the environment-specific characteristics in CSI, instead of learning the complicated channel knowledge in diverse environments. Therefore, a lightweight Transformer network with 2 Transformer blocks are adopted for the reference SAM and the proxy SAM. The $d_\mathrm{model}$ and the dimension of the feed-forward module are 64 and 256, respectively. Post-normalization is employed for SAM, which is empirically for shallower networks and provide better fit and faster convergence.

\section{Simulation Results}
In this section, we first introduce the channel generation settings, training settings, and the baseline methods. Then, we provide numerical analyses among different methods to validate the effectiveness of the proposed methods.

\subsection{Channel Generation Settings}
CSI samples are synthesized using the QuaDRiGa simulator \cite{jaeckel2014quadriga} under the preset \texttt{3GPP\_38.901\_UMi\_LOS} and \texttt{3GPP\_38.901\_UMi\_NLOS} scenario. The BS employs 32 transmit antennas and operates at a carrier frequency of 2.655 GHz, which lies in the 3GPP TS 38.101-1 band ``n7'' \cite{38101}. The system bandwidth is 70 MHz with 32 subcarriers. To create a diverse dataset, we randomly place 110 distinct circular regions inside a cell of radius 200~m. Odd-indexed datasets are non-line-of-sight (NLOS) and even-indexed datasets are line-of-sight (LOS). Each region has a radius of 5~m and contributes a sub-dataset of 10,000 CSI samples. The sub-datasets with indices 1--100 are pooled, randomly permuted, and partitioned into training/validation/test splits with an 8:1:1 ratio for pre-training, while the sub-datasets with indices 101--110 are processed similarly to test the performance of environment adaptation. 

\begin{figure}[t]
    \centering
    \includegraphics[width=0.43\textwidth]{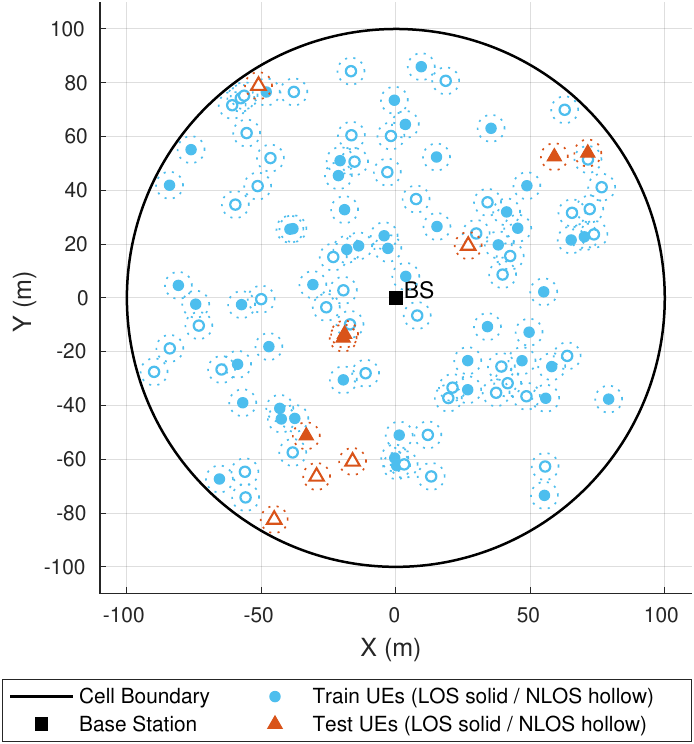}
        \caption{Illustration of CSI data generation scenarios: 110 distinct circular areas (radius 5 m) within a cell (radius 200 m), each generating 10,000 samples.} 
	\label{fig:tradeoff}
\vspace{-0.5cm}
\end{figure}

\subsection{Training Settings}
All neural network models are implemented in TensorFlow 2.14 and trained on NVIDIA TESLA V100 GPUs. We adopt the AdamW optimizer and a batch size of 256. The MSE serves as the training objective, while both Normalized MSE (NMSE) and GCS are reported as evaluation metrics. For the pre-training of the base LAM and the reference SAM, we train each neural network for 500 epochs with 5\% training steps for warm-up, after which a cosine decay learning rate is employed. The initial, peak, and final learning rate are $0$, $10^{-3}$, and $10^{-5}$, respectively. For environment adaptation, we employ a constant learning rate of $10^{-3}$ to better compare the convergence speed of different methods, and early exiting strategy is adopted when the NMSE loss on the validation set is not improved for a consecutive of 20 epochs. The maximum training epochs of environment adaptation is set to 100. The checkpoint achieving the best validation NMSE is retained for final evaluation.

For comparison purpose, we consider following baselines:
\begin{itemize}  
    \item \textbf{Pre-trained LAM:} An LAM identical to the base LAM is pre-trained as described in Step 1, and directly used for  CSI feedback.
    \item \textbf{Pre-trained SAM:} An SAM identical to the reference LAM is pre-trained as described in Step 1, and directly used for CSI feedback.
    \item \textbf{Fine-tuned SAM:}: An SAM identical to the reference LAM is first pre-trained as described in Step 1, and then fine-tuned with the CSI datasets in the target environment.
    \item \textbf{Baseline A:} An LAM identical to the base LAM and an SAM identical to the reference SAM is employed. The output of the LAM and the SAM are added to generate the output reconstructed CSI. The LAM is pre-trained as described in Step 1, then the entirety of the LAM and SAM are fine-tuned with CSI datasets in the target environment with the parameters of LAM frozen.
\end{itemize}

\subsection{Performance Improvement}
\begin{figure*}[t]
    \centering
	\subfigure[]{
		\includegraphics[width=0.45\linewidth]{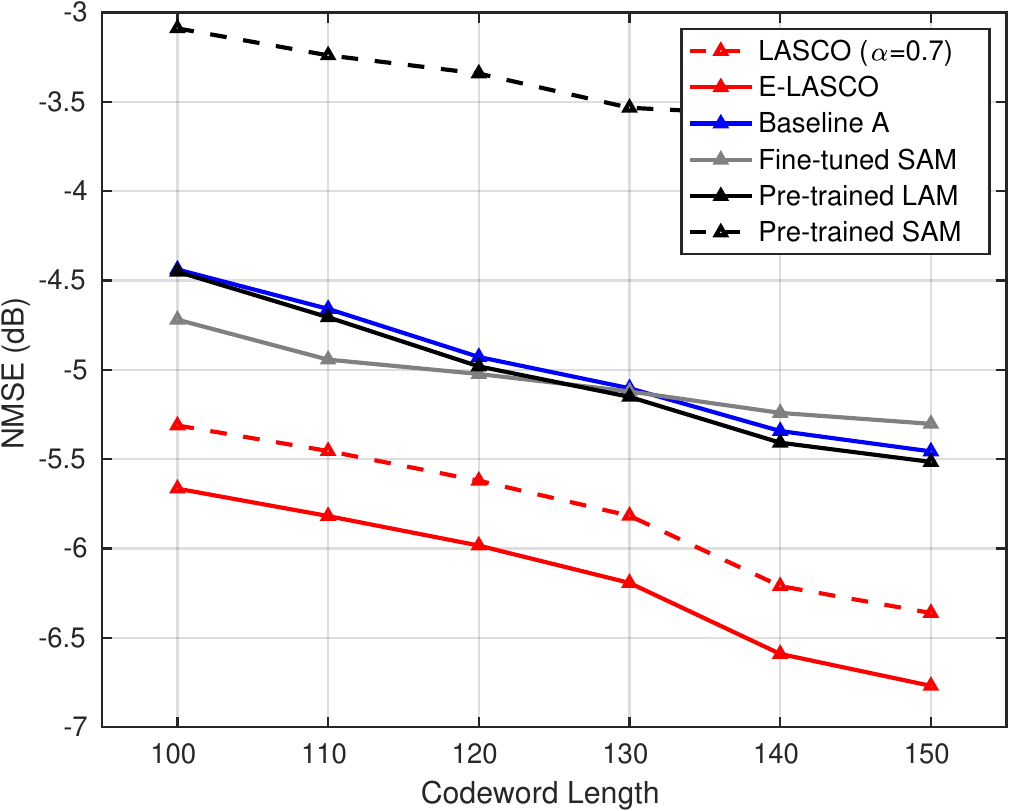}}
        \hspace{0.9cm}
	\subfigure[]{
		\includegraphics[width=0.45\linewidth]{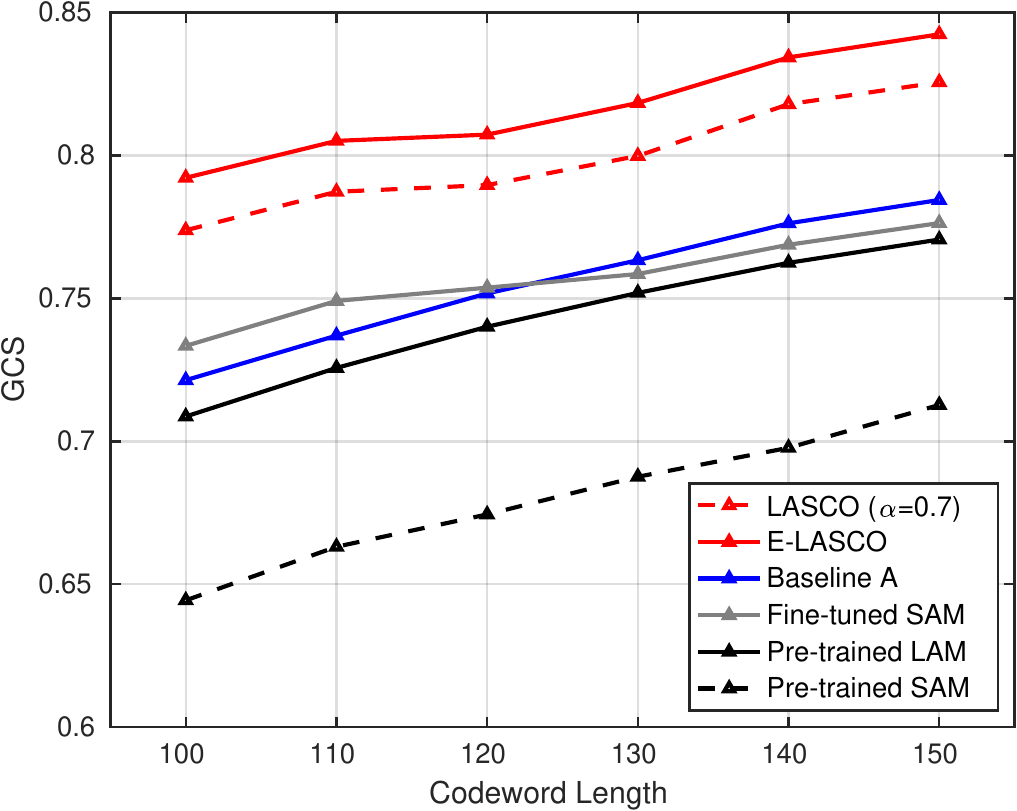}}
        \caption{Performance comparison among different methods under different codeword length. NMSE and GCS are used as evaluation metrics in (a) and (b), respectively. The training samples for each adaptation is set to 8,000.} 
	\label{fig:cr}
     \vspace{-0.5cm}
\end{figure*}

We compare the NMSE performance of LASCO and E-LASCO with baseline methods under different codeword lengths, as illustrated in Fig. \ref{fig:cr}(a). First, the proposed LASCO achieves evident performance gains compared to the pre-trained LAM, indicating that LASCO successfully reaps the benefit of environment-specific adaptation without modifying any parameters of the LAM. Moreover, LASCO outperforms the direct fine-tuning of a pre-trained SAM, implying that it not only realizes the gain of environment adaptation, but also effectively leverages the rich prior knowledge embedded in the LAM for collaborative enhancement of feedback performance. In addition, E-LASCO further improves the performance across different compression ratios by introducing an adaptive collaboration coefficient to elastically fuse the outputs of LAM and SAM. This design enables the system to dynamically adjust the degree of collaboration according to the environment.
For comparison, baseline A concatenates an SAM with the base LAM in parallel and trains the joint network from scratch, but it fails to achieve favorable performance, highlighting the advantage of our collaboration-based design. We also evaluate the GCS performance of all methods in Fig. \ref{fig:cr}(b). The proposed LASCO and E-LASCO still exhibit superior performance compared to other baseline methods, consistently bringing an evident environment adaptation gain under different codeword lengths.

\subsection{Influence of Hyper-parameters}
\begin{figure*}[t]
    \centering
	\subfigure[]{
		\includegraphics[width=0.45\linewidth]{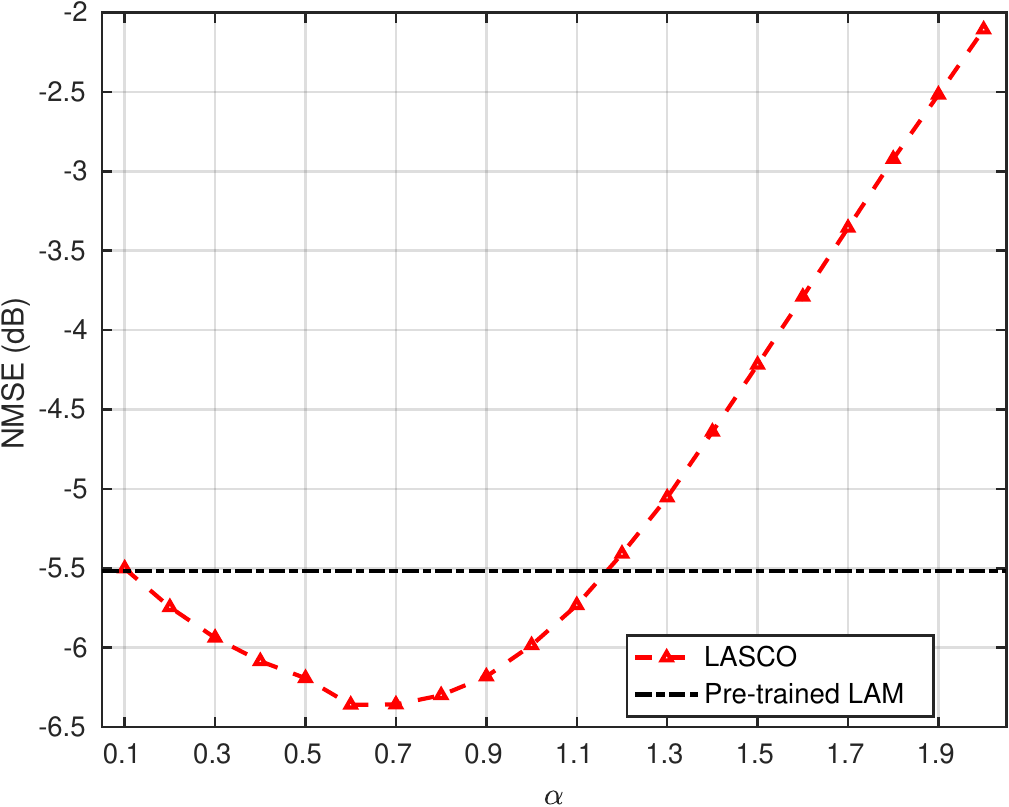}}
        \hspace{0.9cm}
	\subfigure[]{
		\includegraphics[width=0.45\linewidth]{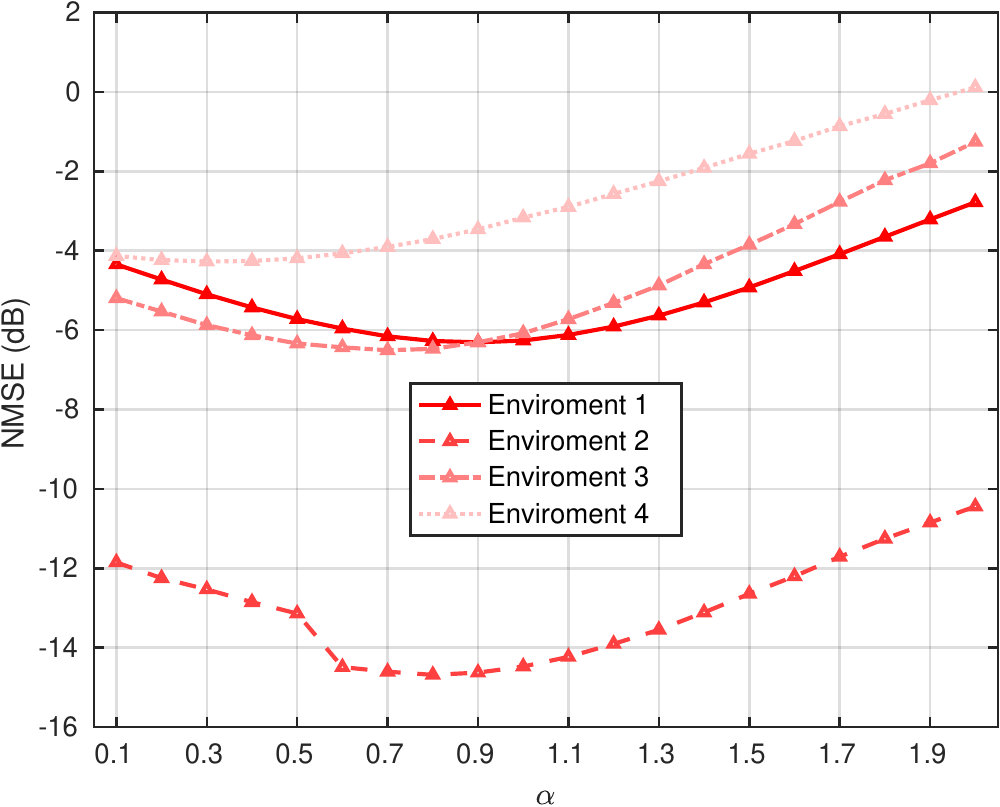}}
        \caption{(a) Average performance of LASCO with different values of the hyper-parameter $\alpha$ when tested on different environments. (b) Performance of LASCO with different hyper-parameters $\alpha$ tested on a single environment. The codeword length and the training samples are set to 150 and 8,000, respectively.} 
	\label{fig:alpha}
    \vspace{-0.5cm}
\end{figure*}

The hyper-parameter $\alpha$ is critical for controlling how the LAM collaborate with the SAM. We test the performance of LASCO with different values of hyper-parameter $\alpha$. First, we compare the average performance of LASCO across different environments. As shown in Fig. \ref{fig:alpha}(a), with increasing $\alpha$ from 0.1 to 2.0, the performance of LASCO first improves and then degrades. When a smaller $\alpha$ is adopted, the base LAM less participates in the collaboration, and LASCO gradually degrades to directly fine-tuning the SAM. As the value of $\alpha$ grows, the base LAM plays a more important role in CSI reconstruction and help provide better performance. When $\alpha$ is set to approximately 0.7, LASCO achieves the optimal average performance. This implys the optimal collaboration between the LAM and the SAMs relies on the effective exploitation of the generalized channel knowledge of the LAM. Moreover, the phenomenon that the optimal $\alpha$ is less than 1 also exhibits an adaptation to the new environment, which is usually accompanied with a proper discard of outdated knowledge. When the $\alpha$ is overlarge, an overstate of the generalized knowledge in the base LAM causes performance degradation of LAM. Therefore, an careful selection of $\alpha$ is critical to the collaboration between LAM and SAMs. 

Then, we conduct per-environment case study to further investigate the influence of $\alpha$ on different environments. As shown in Fig. \ref{fig:alpha}(b), the optimal values of $\alpha$ varies among different environments. For example, LASCO achieves optimal performance when $\alpha$ is around 0.9 for Environment 101, while for Environment 104, the optimal $\alpha$ for LASCO is approximately 0.3. This implies that the LAM and SAMs prefer different collaboration patterns for different environments. While the base LAM can already provide favorable performance for some environments without further adaptation, some other environments more depart from the common environments captured by the base LAM and are more dependent on environment adaptation to capture the distinct channel features. The diversity of the optimal hyper-parameter in different environments further underscores the necessity of elastic design in E-LASCO. When the hyper-parameter is not appropriately set for each single environment, the performance may significantly degrades and cause inefficient collaboration between the LAM and the SAMs.

\subsection{Sample Efficiency in Environment Adaptation}

\begin{figure}[t]
    \centering
    \includegraphics[width=0.45\textwidth]{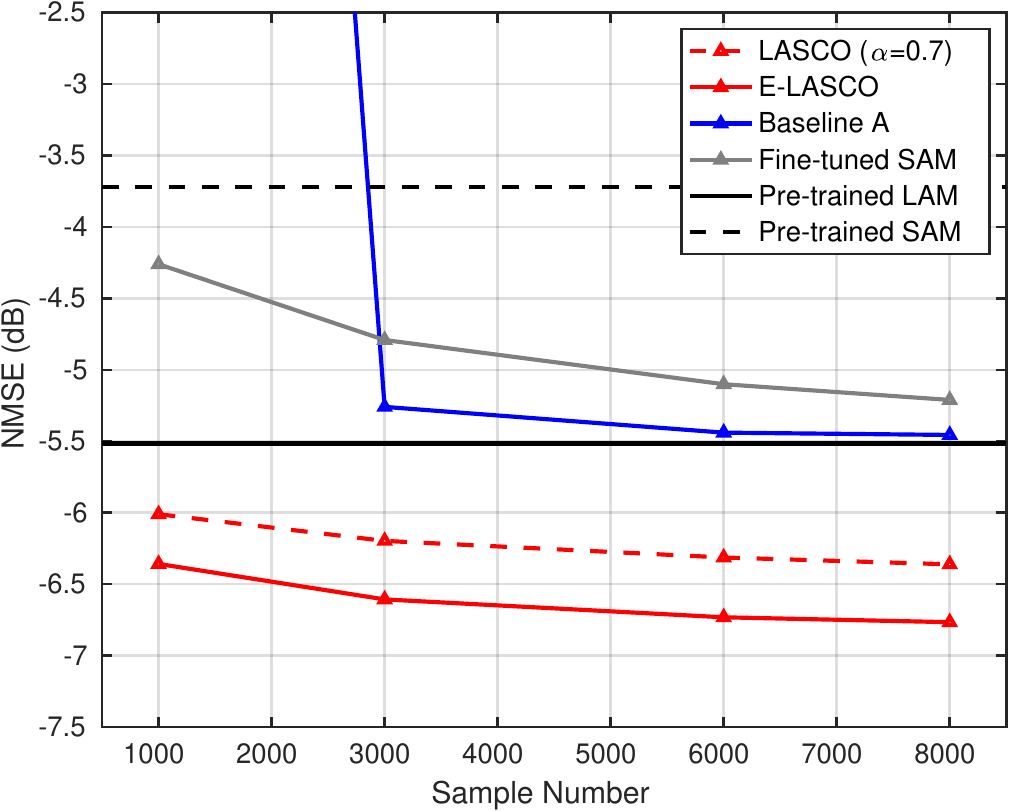}
        \caption{Performance comparison among different methods under numbers of training samples. The codeword length is set to 150.} 
	\label{fig:sample}
\vspace{-0.5cm}
\end{figure}

When adapting to a new environment, CSI data collection might cause significant signaling overhead. Therefore, we further test the performance of different methods when different numbers of training samples are available. As shown in Fig. \ref{fig:sample}, when the training samples are deficient, LASCO still exhibits favorable performance, and continue to benefit from the increase of training samples. In contrast, baseline A fails to converge when only 1000 training samples are available, implying its dependence on a large training datasets. Moreover, directly fine-tuning a pre-trained SAM is relatively more sensitive to the number of training samples, and exhibits an evident performance drop when the number of training samples decreases. The improved sample efficiency in environment adaptation is chiefly due to collaboration with the LAM. Serving as a rich repository of channel knowledge, the LAM offers a strong starting point in new environments and eases the learning burden on the SAM.

\subsection{Why a Reference Model is Needed?}
\begin{figure*}[t]
    \centering
	\subfigure[]{
		\includegraphics[width=0.45\linewidth]{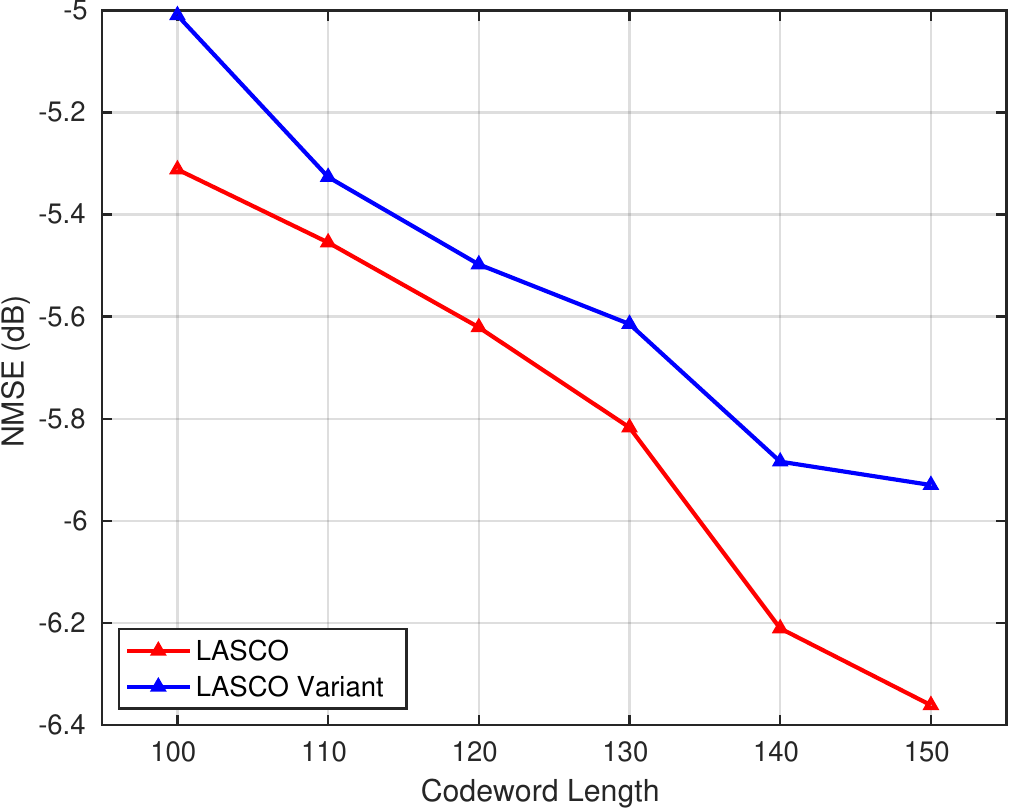}}
        \hspace{0.9cm}
	\subfigure[]{
		\includegraphics[width=0.45\linewidth]{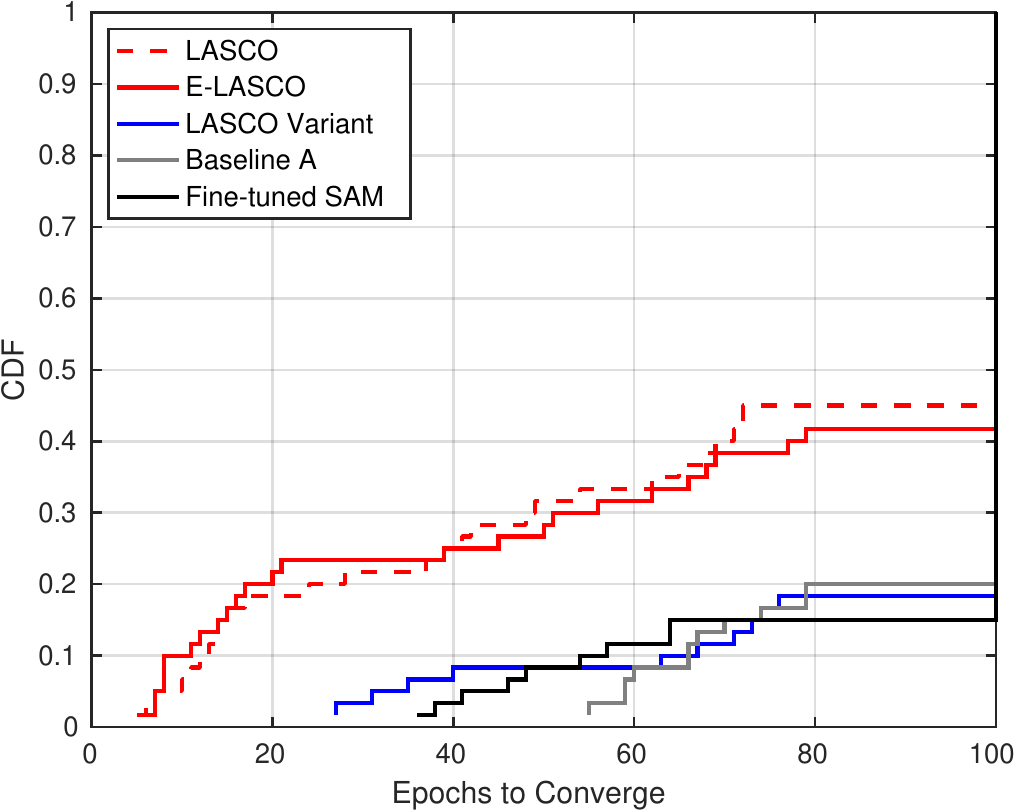}}
        \caption{(a) Performance comparsion between LASCO and variant LASCO, where the training samples for each adaptation is set to 8,000. (b) The CDF of the number of epochs before convergence for different methods, where the codeword length ranges from 100 to 150, and the training samples for each adaptation is 8,000.} 
	\label{fig:zminus}
    \vspace{-0.5cm}
\end{figure*}

Ablation study is conducted to better illustrate the design philosophy of the proposed LASCO framework. We establish a variant of LASCO for comparison purpose, where the reference SAM is discarded. As shown in Fig. \ref{fig:zminus}(a), the variant LASCO exhibits lower performance compared to the original LASCO under different codeword lengths. Also, we compare the convergence speed of variant LASCO and the original LASCO. The training epochs are tested with different environments and different codeword lengths and the cumulative distribution functions of training epochs that different methods need to converge are shown in Fig. \ref{fig:zminus}(b). As shown in Fig. \ref{fig:zminus}(b), in approximately 50\% of test cases, LASCO and E-LASCO requires significantly fewer training epochs to adapt to new environments than the baselines. For another 50\% of test cases, all methods need more than 100 epochs to converge, which is primarily attributed to the difficulty of adaptation tasks. 

According to the results, the reason behind the faster adaptation and the higher performance compared to the variant LASCO is analyzed as follows. The environment adaptation hehavior better benefits from the pre-training with the assistance of the reference SAM. In pre-training stage, the training objective is to let the LAM and SAM reconstruct the CSI as accurate as possible. However, when a reference SAM is absent, the proxy SAM pre-trained to reconstruct CSI is tuned to reconstruct the residual between the ground-truth CSI and the CSI reconstructed by the base LAM, which is deviated from the objective of pre-training. This leads to a larger domain gap and ineffective exploitation of the knowledge acquired in pre-training. In contrast, with the participation of the reference SAM, the objective of the proxy SAM is still to reconstruct the CSI, which better leverages the knowledge acquired from pre-training and results in improved convergence and performance.

\subsection{Influence of Model Size}
When establishing a collaboration framework between LAMs and SAMs, the model size of LAMs and SAMs may influence the collaboration between them. We fix the size of the base LAM and test the SAMs with different model sizes for E-LASCO. As shown in Fig. \ref{fig:tradeoff}, the performance of E-LASCO improves with the increase of the model size of the SAMs. While a larger SAM improves accuracy, it also introduces higher computational cost. Considering both inference complexity and adaptation performance, an appropriate SAM architecture should be carefully selected to provide the best trade-off between accuracy and efficiency.

\begin{figure}[t]
    \centering
    \includegraphics[width=0.45\textwidth]{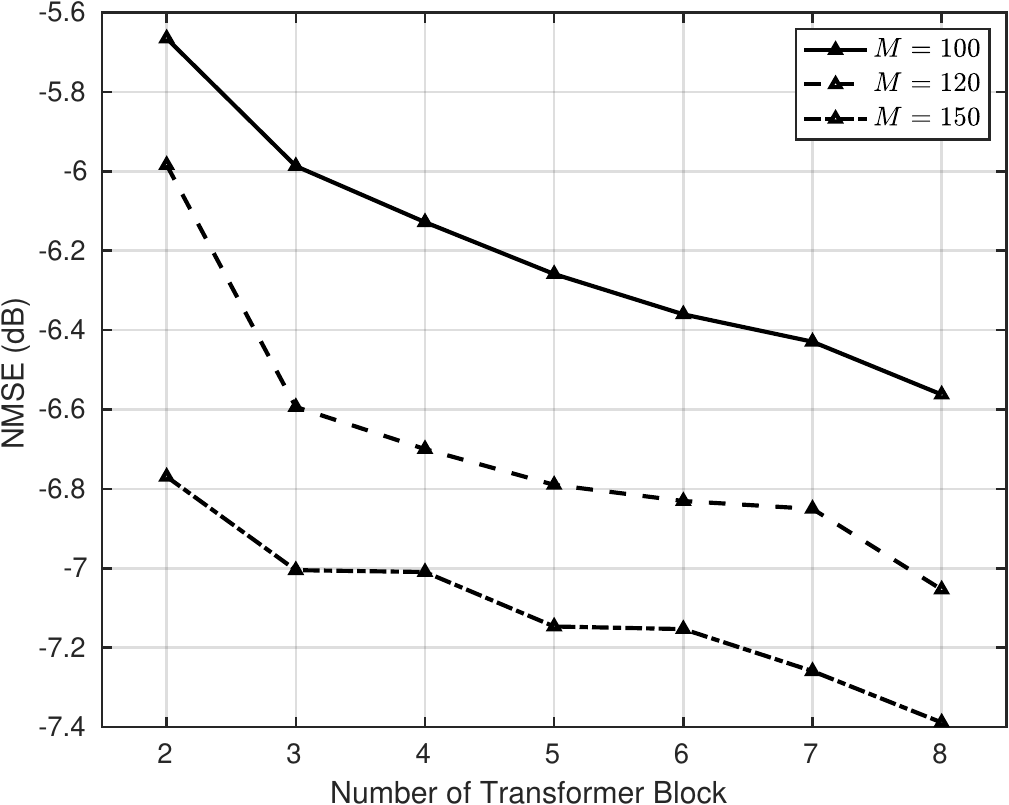}
        \caption{Performance of E-LASCO with different architectures of the reference SAM and proxy SAM under different compression ratios, where the training samples for each adaptation is set to 8,000.} 
	\label{fig:tradeoff}
\vspace{-0.5cm}
\end{figure}

\section{Conclusion}
This paper has presented a comprehensive study on large and small model collaboration for air interface intelligence. We first established a general framework in which LAMs and SAMs play complementary roles: the LAM serves as a foundational channel knowledge base that provides generalized understanding of wireless propagation, while the SAM functions as a lightweight environment-specific plugin that enables rapid adaptation to new deployment scenarios. This framework can be naturally extended to different tasks in air interface. To provide a more concrete and intuitive interpretation of the proposed framework, we developed LASCO, a practical instantiation for CSI feedback, where small models emulate the adaptation behavior of large models by modeling their output distribution shifts across different environments.
To further enhance flexibility, we extended the framework into E-LASCO, in which the collaboration strength between large and small models is learned automatically rather than manually tuned. Extensive simulations verified that the proposed LASCO and E-LASCO achieve environment-specific performance gains without requiring parameter access to large models, offering a data-efficient and computationally lightweight solution for environment adaptation. The results highlight the effectiveness of the large-small collaboration paradigm in bridging the gap between model generalization and environment specificity.
The proposed framework not only provides a viable path for practical deployment of large AI models in wireless systems but also establishes a foundation for future research on intelligent, adaptive, and scalable air interface design.

\bibliographystyle{IEEEtran}
\bibliography{refer1}
\end{document}